# Deprivation, Crime, and Abandonment: Do Other Midwestern Cities Have 'Little Detroits'?

Scott W. Hegerty
Department of Economics
Northeastern Illinois University
Chicago, IL 60625
S-Hegerty@neiu.edu

## ABSTRACT

Both within the United States and worldwide, the city of Detroit has become synonymous with economic decline, depopulation, and crime. Is Detroit's situation unique, or can similar neighborhoods be found elsewhere? This study examines Census block group data, as well as local crime statistics for 2014, for a set of five Midwestern cities. Roughly three percent of Chicago's and Milwaukee's block groups—all of which are in majority nonwhite areas—exceed Detroit's median values for certain crimes, vacancies, and a poverty measure. This figure rises to 11 percent for St. Louis, while Minneapolis has only a single "Detroit-like" block group. Detroit's selected areas are more likely to be similar to the entire city itself, both spatially and statistically, while these types of neighborhoods for highly concentrated "pockets" of poverty elsewhere. Development programs that are targeted in one city, therefore, must take these differences into account and should be targeted to appropriate neighborhoods.

# I. Introduction

While many of Northern U.S. cities have experienced economic decline and steep population declines over the past half-century, nowhere are these effects more prominent than in the city of Detroit. Having lost roughly two-thirds of its residents since its 1950s peak of two million, Detroit is now known worldwide for its vacant buildings and inability to fund basic services such as adequate street lighting. While this city serves as an extreme case, it often evokes comparisons with other U.S. cities. These might be overgeneralized by area residents and applied to all urban areas (often negatively), or by government officials who seek to apply other cities' best urban-development strategies—and avoid ones that have failed elsewhere.

Are other U.S. cities really comparable to Detroit? Are there pockets that are similar, are comparisons unwarranted outside of this extreme case? While Detroit is the subject of numerous studies, to our knowledge, no study has directly compared neighborhoods across cities in this way. Smith (2009), for example, discusses Detroit and Portland (Ore.), but in the context of racial characteristics and urban pollution. This study attempts to establish a set of criteria by which U.S. Census block groups can be compared to the city of Detroit. Once they are isolated, these "Little Detroits" can be mapped and examined for socioeconomic characteristics. Of the four Midwestern cities studied here, St. Louis has the highest proportion of such areas, and Minneapolis has the fewest.

# II. More Than Economic Deprivation: Finding Suitable Measures

Our starting point for this analysis involves calculating economic deprivation, which captures not only poverty, but other variables into a single multivariate index. Pioneered by Townsend (1987), who used 77 indicators (including diet, clothing, housing, education, and social inclusion) to measure disadvantage in Britain, this approach has since included a smaller set of variables. Salmond *et al.* (1998), for example, conduct principal component analysis (PCA) to capture the common variance in a set of ten variables (including the receipt of benefits, the unemployment rate, schooling, occupancy, and housing tenure) in New Zealand. A number of comparisons among measures have been performed, by Morris and Carstairs (1991), Carstairs (1995), and others.

These measures are often subject to criticism. Pacione (2002) finds that techniques used for urban Britain are not applicable to rural Scotland, while Deas and Robson (2003) call for the inclusion of crime, air quality, and other measures of environmental degradation in Britain's indices. In a critical review of the literature, Fu *et al.* (2015) note that census-based methods may be biased include due to the inclusion of age- and gender-based variables (such as the proportion of female-headed households) in many deprivation measures. Because existing methods are not tailored to conditions in Detroit and cross-city comparisons, we therefore take a different approach.

As we note below, creating a single deprivation index that can be compared with sufficient precision across multiple cities can be problematic. Data come from different sources, and out of a number of potential weighting schemes for a multivariate index, some might not allow for direct comparisons. We also wish to focus on crime not only as a threat to public safety, but also as a measure of neighborhood conditions. As Sampson (1985), Peterson and Krivo (2010), and Raleigh and Galster (2015) note, deprivation helps contribute to reduced institutional capacity, which, as Social Disorganization theory suggests, leads to increased crime rates. Finally, while vacancy rates are often included in deprivation indices, we include this variable as a separate component of our criterion due to the sheer importance of this issue for the city of Detroit.

Rather than relying on a single deprivation index, we instead focus on one major (unweighted) aspect of deprivation, as well as separate measures of crime and vacancies to create a trivariate criterion for neighborhoods that exceed the Detroit median values. Areas that meet this criterion, in the Midwestern cities of Chicago, Milwaukee, Minneapolis, and St. Louis, are then mapped and statistically examined to uncover their underlying income, racial, and socioeconomic characteristics.

### III. Methodology

In this study, we use three concepts to measure urban characteristics: poverty or economic deprivation, crime, and vacancies. We construct our measure for U.S. Census block groups for Detroit, Chicago, Milwaukee, Minneapolis, and St. Louis for the year 2014. Various socioeconomic and housing variables are available from the U.S. Census (2014 ACS 5-year estimates). We also obtain address data from individual city police department websites for reported crimes, which we geocode and apply to the relevant block groups. Individual variables are described below. Finding a common measure for these three variables across all cities presents a key challenge, however.

As a preliminary step, we must choose an appropriate measure of deprivation. We begin by calculating a four-variable index that includes the unemployment rate, the percentage of adults over age 25 without a high-school diploma, the percentage of SNAP recipients[1], and the percentage in poverty. The weighting scheme for this index can be problematic, however. While we prefer using principal Components Analysis (PCA) to isolate the common variance among the four variables, doing so for each city would make direct comparisons difficult. Rather than pooling all four cities' block groups for a single PCA deprivation measure, we employ a widely used alternative measure, weighting each variable by its standard deviation. We construct two separate measures of this type, one that uses each city's own standard deviations over all block groups, and another that uses Detroit's standard deviations for all cities, as follows:

$$DEP = \frac{1}{\sigma_1} POV + \frac{1}{\sigma_2} SNAP + \frac{1}{\sigma_3} UNEMP + \frac{1}{\sigma_4} NOHS \tag{1}$$

---

[1] Supplemental Nutrition Assistance Program, a form of public assistance.

We therefore calculate three separate deprivation measures, which we show below to be highly correlated with one another. We prefer the "Detroit weights" deprivation index to the "Own weights" and the "PCA" indices because of their consistency across cities. But, since we find that the proportion of SNAP recipients is the index component that is most highly correlated with all deprivation indices, we choose to consider this variable, unweighted, when forming our trivariate criterion.

Secondly, crimes might be reported or classified differently in different jurisdictions. In fact, not every U.S. city publishes an entire year's historical crime data, thereby restricting the number of candidates for this type of study. For those that do provide adequate data, police or data-collection resources might be more limited in some cities, crime categories might be defined differently, or or established practices can vary from city to city. As an example, Milwaukee does not provide address data for sexual crimes, so any aggregate values would not match those of cities that do.

To maximize the uniformity of crime categories, we focus on a subset of crimes: Assaults, Burglaries, and Robberies. These are hopefully easy to compare across jurisdictions not only because the categories are clear, but also because they capture a level of "invasiveness" that might lead to increased fear among residents. While Homicides are high-profile, they are also much less common, and are often specific to the victim rather than a neighborhood. These three crimes, therefore, comprise a sufficient measure of neighborhood conditions that can be compared across the five cities of our study.

Finally, we must address the fact that census-defined "vacancies" might not be equivalent to urban-prairie-type areas where homes have been previously demolished, leaving no more buildings to be unoccupied. Ideally, a measure such as those constructed using Detroit's vacancy index database would capture true levels of abandonment. But, not all cities conduct detailed surveys, and these cannot be expected to be consistent across cities. We therefore use Census data to compare the percentage of housing units vacant for all cities. As we show below, this variable is sufficiently correlated with values from by comparing Detroit's detailed survey.

Having chosen our three measures—percentage of SNAP recipients; Assaults, Burglaries, and Robberies (*ABR*) per capita; and the vacancy rate—we then must choose how to define "Detroit-type" block groups. We do this with a simple technique: choosing block groups that exceed Detroit's median values for all three measures. It is important to note that including crime and vacancies separately from deprivation (and not examining a single, six-variable index) ensures that all three criteria are met. In an index, crimes and other variables could be substitutable; for example, a block group with excessive unemployment but low crime could have a higher index value than the type of area that we are seeking to define as the object of this study.

Examining the proportion of such poor, high-crime, and high-vacancy areas for each comparison city helps show how the prevalence of urban decline differs. Within each city, mapping them can help identify the neighborhoods with the greatest need. As we show later in this paper, these block groups are nearby, but not identical to, each city's highest-deprivation areas. This helps establish the fact that these areas are prone to more than simple economic deprivation.

We then examine how a set of socioeconomic variables differ within these selected areas compared to each city as a whole. Many differences are not large. We calculate Z-scores to show the number of (whole-sample) standard deviations by which the selected areas' means differ from the citywide mean. While we do not make any assessments based on standard significance levels, we are able to see the variables with the largest differences across groups.

Finally, we consider whether our results are similar if only economic deprivation is considered. We map the block groups with the highest deprivation industries for each city, choosing the number of Detroit-like areas as our cutoff. We can also compare whether socioeconomic variables differ depending on our choice of measurement. Our results are provided below.

## IV. Results

The five cities examined in this study differ greatly in terms of their relative level of economic and social well-being. Figure 1 provides simple summary statistics for the major variables in our analysis: The four deprivation components and our three alternative measures, as well as *ABR* (crimes) per capita, and the vacancy rate. Detroit has the largest average values of all variables except the crime rate, where it is second to St. Louis. While it is possible that this might be due to differences in crime reporting, we assume the proportions to be as accurate as possible. Minneapolis consistently has the lowest average values. When comparing these ranks, we find that St. Louis' consistent second- or third-place rankings puts it closer to Detroit, with Milwaukee and Chicago roughly comparable to one another.

The standard deviations help us show how different weighting schemes might lead to differences in the construction of deprivation measures across cities. The poverty has the highest variance (and thus the smallest weight in our chosen index) for all cities except Chicago. The percentage of adults over 25 with no high-school diploma would have the largest weight in individual indices for Detroit and St. Louis, while the vacancy rate has the smallest standard deviation for the other cities. Differences also arise if we compute factor loadings using PCA. While we also construct PCA and "own weight" deprivation indices, we weight all five cities' deprivation indices using Detroit's standard deviations. Table 2 shows all three indices to be highly correlated with one another, and also correlated with their component variables. The percentage of SNAP recipients is the most highly correlated with deprivation for all cities.

We map deprivation for all five cities in Figure 1. The shading classifications represent Detroit's natural series breaks, and are applied to the other four cities as well. If colors are lighter overall, (as in the case in Minneapolis), this indicates that the relevant values are lower compared to Detroit. Deprivation is clearly more extreme, and less concentrated, in Detroit. Deprivation is lower outside Northwest and Southeast Minneapolis, while many of the northern areas of St. Louis have deprivation levels that are comparable to Detroit's. The South and West sides of Chicago, as well as the near North side of Milwaukee, also have concentrations of economic deprivation.

The maps suggest that deprivation is fairly equally spread throughout Detroit, but not the other cities. This is indeed confirmed by the Moran's *I* coefficients in Table 3, which are significantly higher for the other four cities in this study. The same is true for the percentage of SNAP recipients. The other two variables included in the table show some different results, however: Crime has the lowest coefficient in Chicago, and vacancies in Minneapolis are roughly equally autocorrelated in all cities except Minneapolis.

As we note above, Detroit's level of abandonment goes far beyond merely empty buildings, yet not all cities catalog their abandoned properties as extensively. We therefore use the Census vacancy rate for block groups in all five cities. We compare Detroit's Census-based vacancy rate to Detroit's vacancy index's "Likely Vacant" and "Very Likely Vacant" categories. Figure 2 confirms that at the census-tract level, average vacancy rates are similar visually, and the correlations presented in Table 3 are high enough to justify the use of a variable that is consistent across cities.

We next look at the spatial patterns of our three measures of urban conditions, which are shown in Figure 3. In general, the percentage of SNAP recipients, per-capita crime rates, and vacancies match each city's most deprived areas. Detroit, again, stands out for exhibiting inner-city" characteristics throughout the city, rather than in isolated areas. We suspect that our selected Detroit-like areas will match these areas.

These areas are mapped in Figure 4, along with the nonwhite share of each city's population. While we did not include any racial characteristics in our selection criterion, it is clear that, much like Detroit itself, these areas in all our cities are primarily African-American. Not all of Detroit's 868 block groups are simultaneously above the citywide median for all three variables, but the areas that are, again, are distributed throughout the city. Minneapolis, the wealthiest city in our sample, has only a single block group that fits our criteria.

Table 5 provides the proportion of block groups that can be described as "Detroit-like" for five cities, as well as a description of how a number of socioeconomic variables differ between these areas and the city as a whole. Detroit (with 18 percent of the total) has the highest proportion that fits our criteria, followed by St. Louis (10.9 percent), Chicago (3.3 percent) and Milwaukee (2.9 percent). We also compare the median values of a number of important socioeconomic variables within our selected areas and within each city. Besides our deprivation index (*SD4DET*) and the three variables used to select these areas (*PERCSNAP, PERCVAC,* and *ABRPOP)*, we also consider the following variables:

*MEDVAL* = the median housing value
*PERCHOUS30K* = The percentage of houses with values below $30,000
*PERCBLACK* = The percentage of the population that is black
*PERCWHITE* =The percentage of the population that is white
*PERCPUBTRA* = The percentage of the population that commutes via public transport
*REVCOMM* = The proportion of reverse commuters with jobs outside the city proper
*PERC40COMM* = The percentage of the population with commute times larger than 40 minutes
*PERCRENT* = The percentage of the population that rents rather than owns

How similar are the selected areas of other cities to the corresponding areas of Detroit? Some differences between Detroit and other cities are particular to regional and citywide economic disparities, but there are valuable comparisons to be made. Chicago is a case in point. The largest city in the U.S. Midwest has higher housing values and a denser public transit network than many of its neighbors in the region. For those reasons, then, it is not surprising that Chicago has higher *MEDVAL*, *PERCHOUS30K*, and *PERCPUBTRA*, even in the selected, "Detroit-like" block groups. But some variables can be more directly compared. In particular, economic deprivation (*SD4DET*) has nearly the same average value in the selected areas of Detroit and Chicago, as are *PERCBLACK* (more than 95 percent for both) and the percentage of SNAP recipients (more than half). The same is true for Milwaukee and St. Louis, with an average deprivation index above 9, more than 50 percent SNAP recipients, and more than 90 percent African-American residents in its selected block groups.

Comparing the selected areas of each city with the entire city as a whole, we find that in general, the selected block groups have lower income, median home values, reverse commuters, and percentages of white residents, and higher percentages of workers with long commutes,

African-American renters, and public transit users. But one key finding is that for Detroit, many selected-area median values are similar to the citywide averages. The median block group is roughly 96 percent African American for both, for example. The comparison cities have much larger disparities.

To help understand these divergences, we calculate Z-scores $\frac{(\bar{X}_{Sel} - \bar{X}_{city})}{\sigma_{city}}$ using city and selection means and the city's standard deviation for each city and variable. While we do not focus on any level of statistical significance, we can note where the differences are largest. These statistics are reported in Table 6.

Detroit's level of deprivation (*SD4DET*) in the selected areas is similar to citywide levels, confirming our conjecture that Detroit does not contain specific "pockets" of concentrated poverty. The three comparison cities do, however, with deprivation levels in the selected areas more than one standard deviation above the city mean. For no variable does the mean in Detroit's selected areas exceed the citywide variance by more than 0.2 standard deviations, except for crime (*ABRPOP*) and the percent white. The small difference in median values for this latter variable, shown above, suggest that this is due to a small variance. Large Z-scores are shown for Milwaukee's crime rate and the percent black in Chicago and Milwaukee (St. Louis' score is close to 1 as well). The percentage of SNAP recipients and the vacancy rate are also much higher in the selected areas for Chicago, Milwaukee, and St. Louis, justifying their inclusion in our selection criterion.

As a robustness check, we compare purely deprived areas with those chosen via our selection criterion. We choose two cutoff points, first taking the 10 percent of each city's block groups with the highest *SD4DET* values, and next taking the same number of block groups as met our original selection criterion. Statistical properties for each subset are provided in Table 7, and the selected and most-deprived block groups are mapped side-by-side in Figure 7.

We see that the smaller groups of deprived areas are located in the same general areas of each city as the selected areas, but that the specific block groups are not identical. The most significant discrepancies can be found on Chicago's Uptown neighborhood and Milwaukee's near South Side, where highly deprived areas that do not meet our trivariate selection criterion are located. This corroborates our idea that these "Detroit-like" neighborhoods are not simply the most deprived ones. While median crime rates are similar between the larger and smaller subsets, the smaller groups are expected to be poorer and have lower property values.

Table 8 compares median values between the selected "Detroit-like" areas and the small-group, highly deprived areas within Detroit and our three comparison cities. Median deprivation is higher, and crime and vacancy rates lower, in the deprivation-only measure; this is expected since our selection criterion is more restrictive. The most deprived block groups have higher median values of renters, reverse commuters, and SNAP recipients. The "Detroit-like" areas are whiter in all cities except St. Louis, and have higher median home values in Milwaukee and St. Louis. Overall, we can conclude that our selection criterion identifies areas that are plagued by more than simply poverty.

### V. Conclusion

With a combination of blight, crime, and poverty, the city of Detroit is considered worldwide to be an "extreme case" of urban decline. At the same time, other cities have drawn comparisons to this extreme case—often critical, but also when urban policies are compared. How unique, exactly, is the city of Detroit in the U.S. context? This study seeks to answer this question using quantitative measures for five cities' block groups for the year 2014.

First calculating a measure of economic deprivation, we find that this measure not only does not capture all appropriate characteristics, it also might not be entirely comparable from city to city. We therefore calculate Detroit's levels of SNAP recipients, vacancy rates, and certain crimes per capita. We compare these values for four reference cities, identifying those block groups that exceed Detroit's median values for all three variables. Of our four cities, St. Louis has the highest proportion of block groups that meet this criterion, while Minneapolis has the lowest. In between, Milwaukee and Chicago have "Detroit-like" areas, which are concentrated in the inner city. Interestingly, while fewer than half of Detroit's block groups meet our trivariate criterion, the selected areas are much more dispersed throughout the city. But, selected block groups in Chicago, Milwaukee, and St. Louis are often statistically similar to corresponding block groups in Detroit. This suggests that these block groups—but not entire cities—can indeed be directly compared to parts of Detroit.

We then examine the statistical properties of the selected areas and compare them with each city as a whole. We find that Detroit's most extreme block groups are not much different from the city itself, while the other cities' selected areas are poor and less white, and have lower home values, than the city proper. This confirms the idea that, while many U.S. cities suffer from urban problems, direct comparisons to Detroit are not appropriate. Certain areas may indeed benefit from policies that are tailored to extreme cases, but many other neighborhoods, which are much less deprived, would require a completely different set of policy tools.

**Table 1: Deprivation Indices, 2014**

*Detroit*

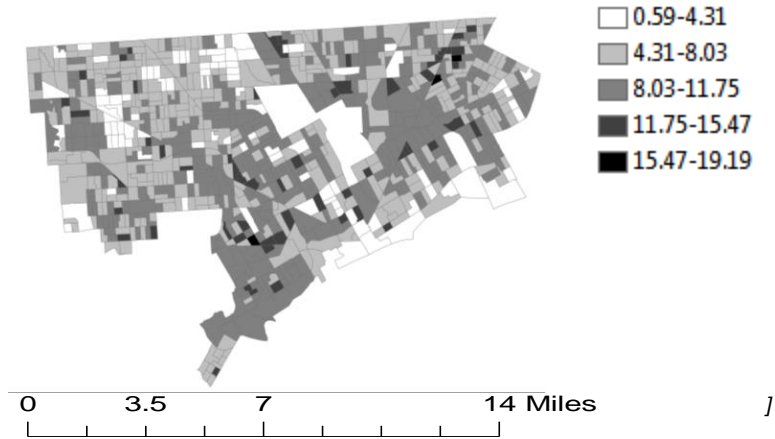

*Chicago*  *Milwaukee*  *Minneapolis*  *St. Louis*

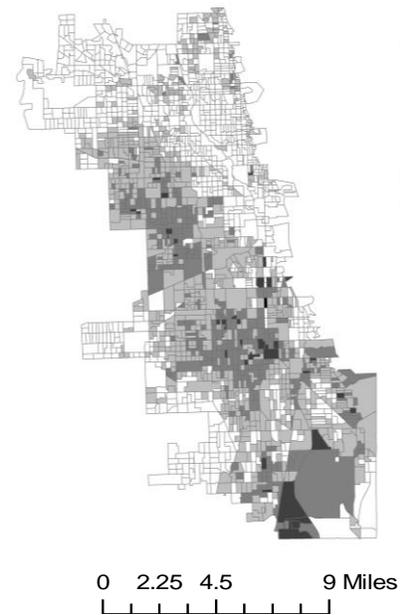 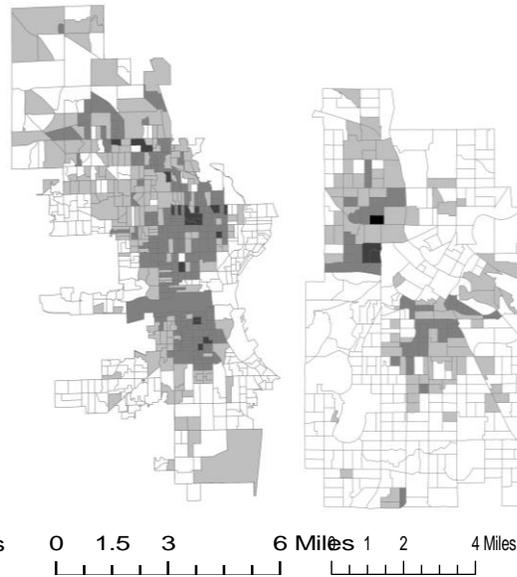 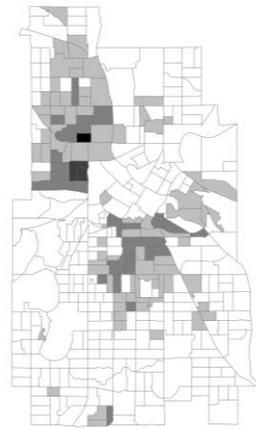 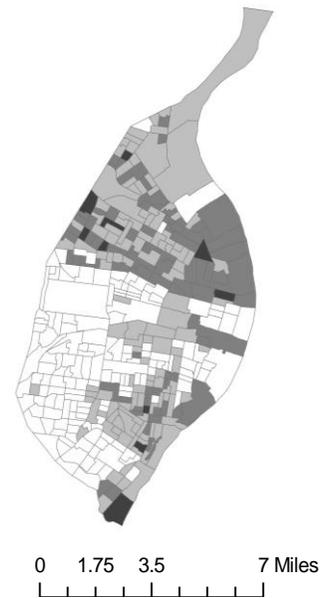

**Table 1. Summary Statistics for Major Variables.**

|  | Detroit Mean | SD | Chicago Mean | SD | Milwaukee Mean | SD | Minneapolis Mean | SD | St. Louis Mean | SD |
|---|---|---|---|---|---|---|---|---|---|---|
| ABRPOP | 0.070 | 0.045 | 0.044 | 0.163 | 0.044 | 0.038 | 0.015 | 0.013 | 0.159 | 0.125 |
| SD4OWN | 7.796 | 2.711 | 4.769 | 3.097 | 5.301 | 3.496 | 3.392 | 3.058 | 5.373 | 3.299 |
| SD4DET |  |  | 3.516 | 3.468 | 5.240 | 3.382 | 3.778 | 3.258 | 5.420 | 3.265 |
| PCA4 | 0.000 | 1.375 | 0.000 | 1.574 | 0.000 | 1.698 | 0.000 | 1.650 | 0.001 | 1.634 |
| PERCNOHS | 22.919 | 13.469 | 18.909 | 14.743 | 17.735 | 14.159 | 10.964 | 12.300 | 17.806 | 11.601 |
| PERCPOV | 36.096 | 20.205 | 19.036 | 17.208 | 22.934 | 19.467 | 16.495 | 20.408 | 22.404 | 19.838 |
| PERCSNAP | 43.451 | 16.762 | 22.260 | 18.549 | 29.111 | 20.824 | 15.815 | 17.168 | 28.359 | 19.507 |
| PERCVAC | 30.973 | 15.269 | 13.861 | 11.145 | 10.423 | 9.456 | 7.614 | 7.284 | 20.788 | 14.526 |
| UNEMP | 15.106 | 8.780 | 9.342 | 6.728 | 8.340 | 5.892 | 6.616 | 5.674 | 9.787 | 7.438 |

**Table 2. Correlations Between Chosen Deprivation Measure and Other Variables.**

|  | PCA4 | SD4OWN | PERCNOHS | PERCPOV | PERCSNAP | UNEMP |
|---|---|---|---|---|---|---|
| Detroit | 0.994 |  | 0.588 | 0.751 | 0.787 | 0.587 |
| CHI | 0.938 | 0.996 | 0.681 | 0.848 | 0.879 | 0.718 |
| MKE | 0.998 | 0.996 | 0.810 | 0.881 | 0.957 | 0.722 |
| MPLS | 0.998 | 0.996 | 0.810 | 0.846 | 0.936 | 0.684 |
| STL | 0.997 | 0.995 | 0.756 | 0.838 | 0.926 | 0.721 |

**Table 3. Moran's I for Spatial Autocorrelation.**

| City | SD4DET | PercSNAP | ABR | VAC |
|---|---|---|---|---|
| Detroit | 0.258 | 0.206 | 0.110 | 0.411 |
| Chicago | 0.693 | 0.654 | 0.048 | 0.395 |
| Milwaukee | 0.681 | 0.633 | 0.484 | 0.420 |
| Minneapolis | 0.621 | 0.591 | 0.428 | 0.123 |
| St. Louis | 0.552 | 0.545 | 0.379 | 0.413 |

**Figure 2. Census Tract-Level Indices of Vacancies, Detroit, 2014.**

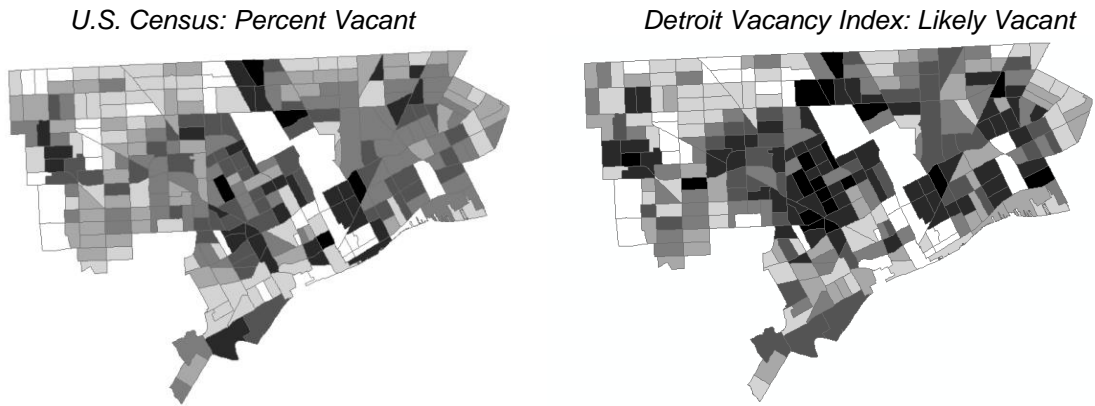

*U.S. Census: Percent Vacant*   *Detroit Vacancy Index: Likely Vacant*

**Table 4. Correlations between Census Data (Percent Vacant) and Detroit Vacancy Index**

|  | Likely | Very Likely | Likely + Very Likely |
|---|---|---|---|
| Pearson | 0.728 | 0.752 | 0.798 |
| Spearman | 0.758 | 0.765 | 0.981 |
| N=293 |  |  |  |

**Figure 3: Block-Group Measures of Socioeconomic Indicators.**

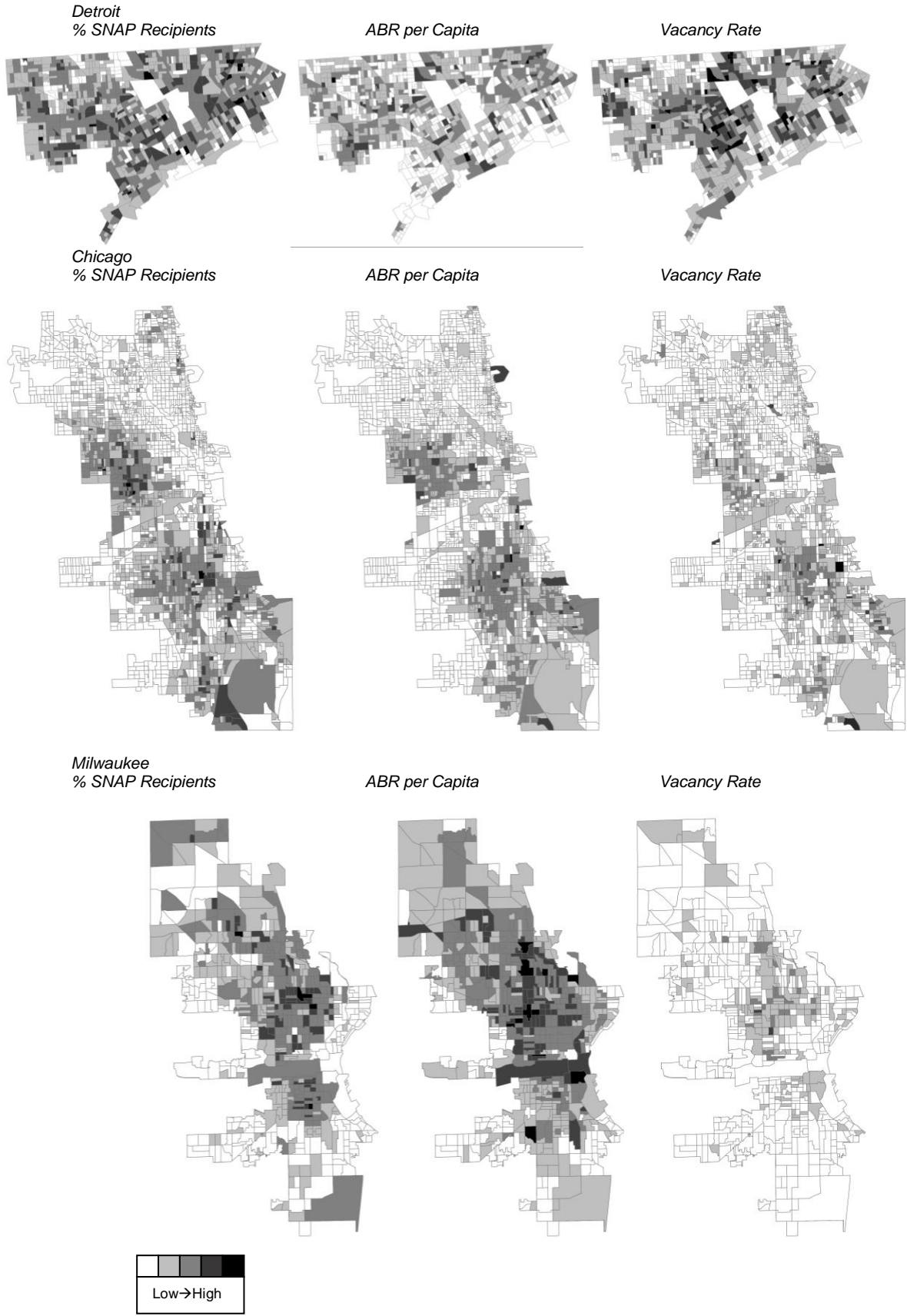

*Minneapolis*
*% SNAP Recipients*   *ABR per Capita*   *Vacancy Rate*

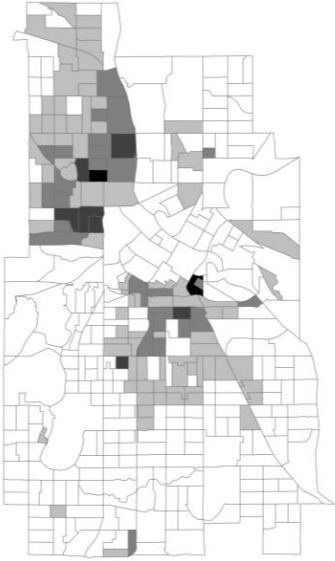 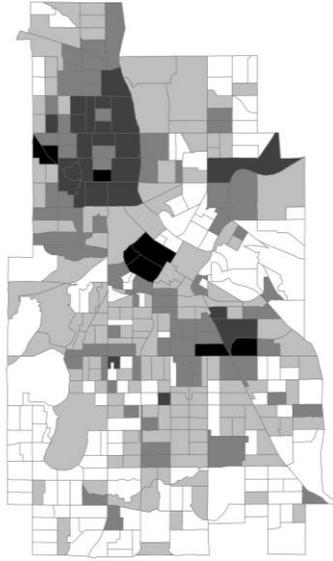 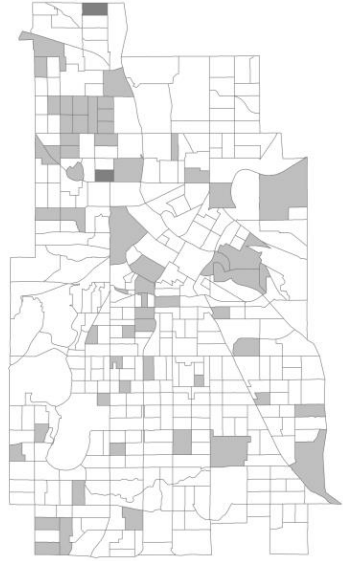

*St. Louis*
*% SNAP Recipients*   *ABR per Capita*   *Vacancy Rate*

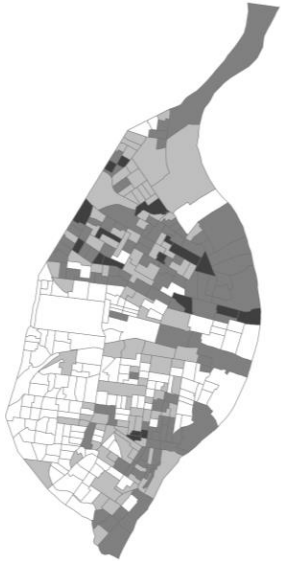 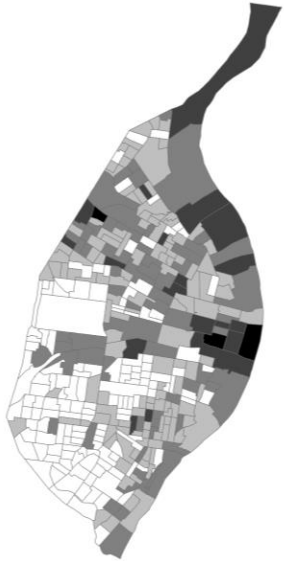 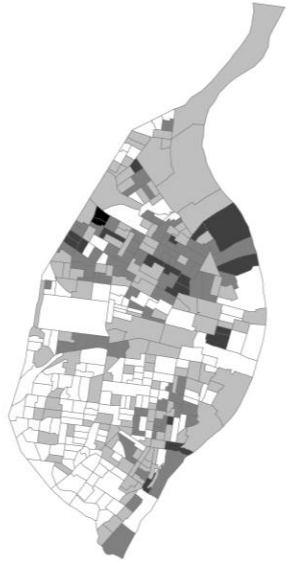

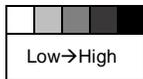
Low→High

**Figure 4. Selected Deprived Areas (Orange) and Percent White (Shaded)**

*Detroit*

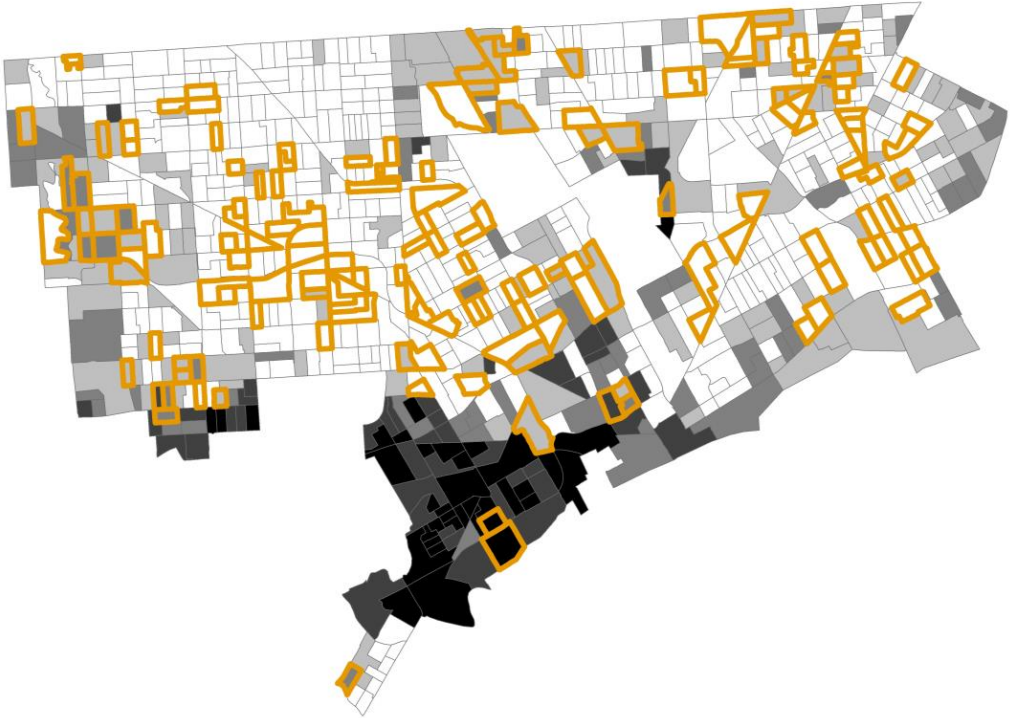

*Minneapolis*

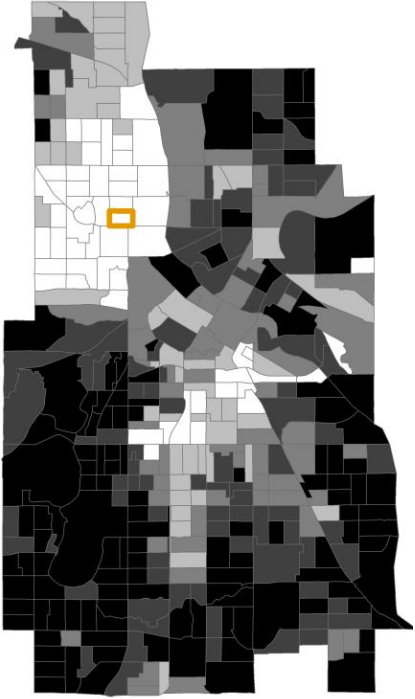

*St. Louis*

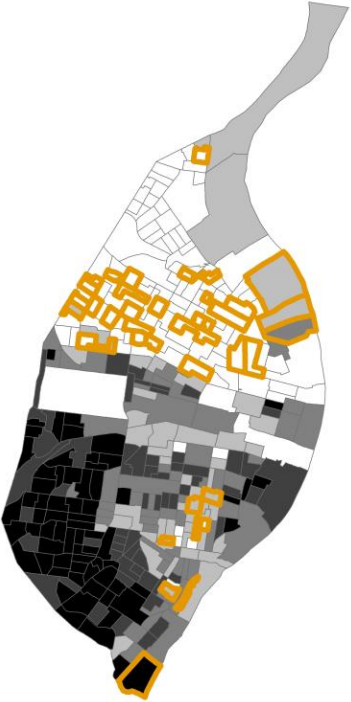

*Chicago*  *Milwaukee*

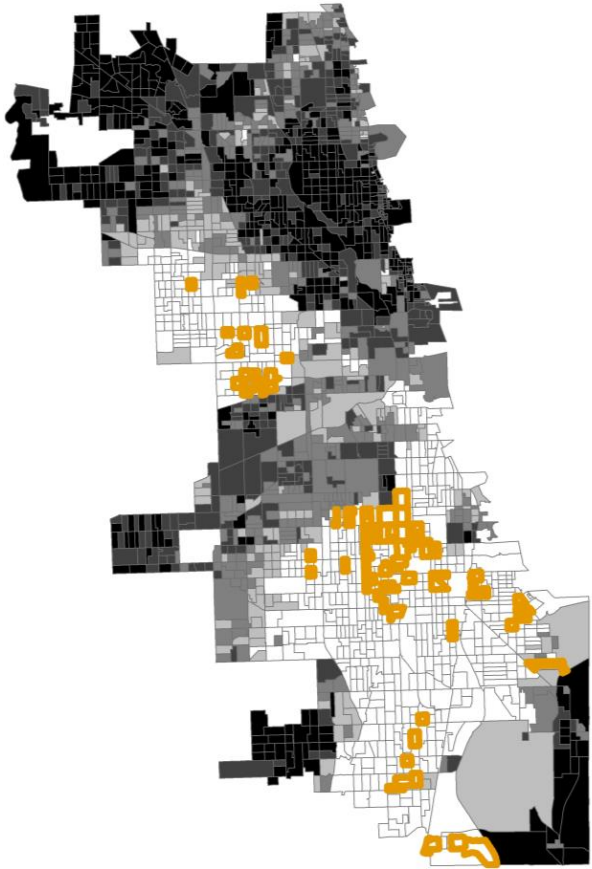
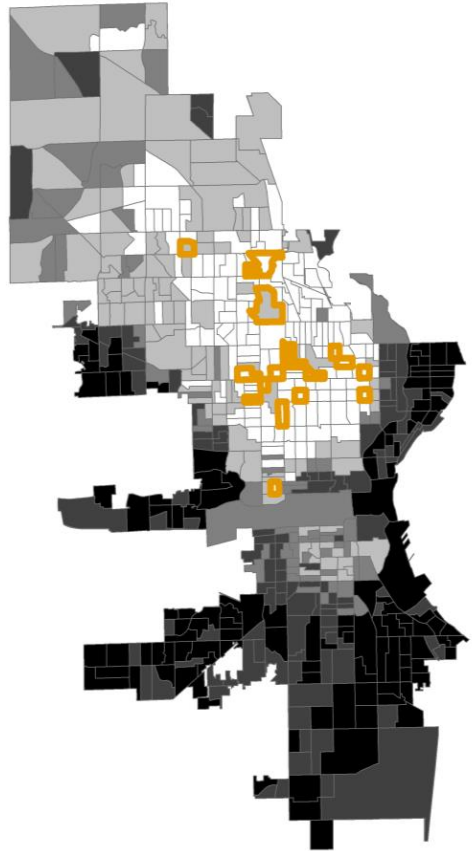

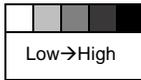
Low→High

**Figure 5. Selected "Detroit-like" Areas and Most Deprived Areas (By Index)**

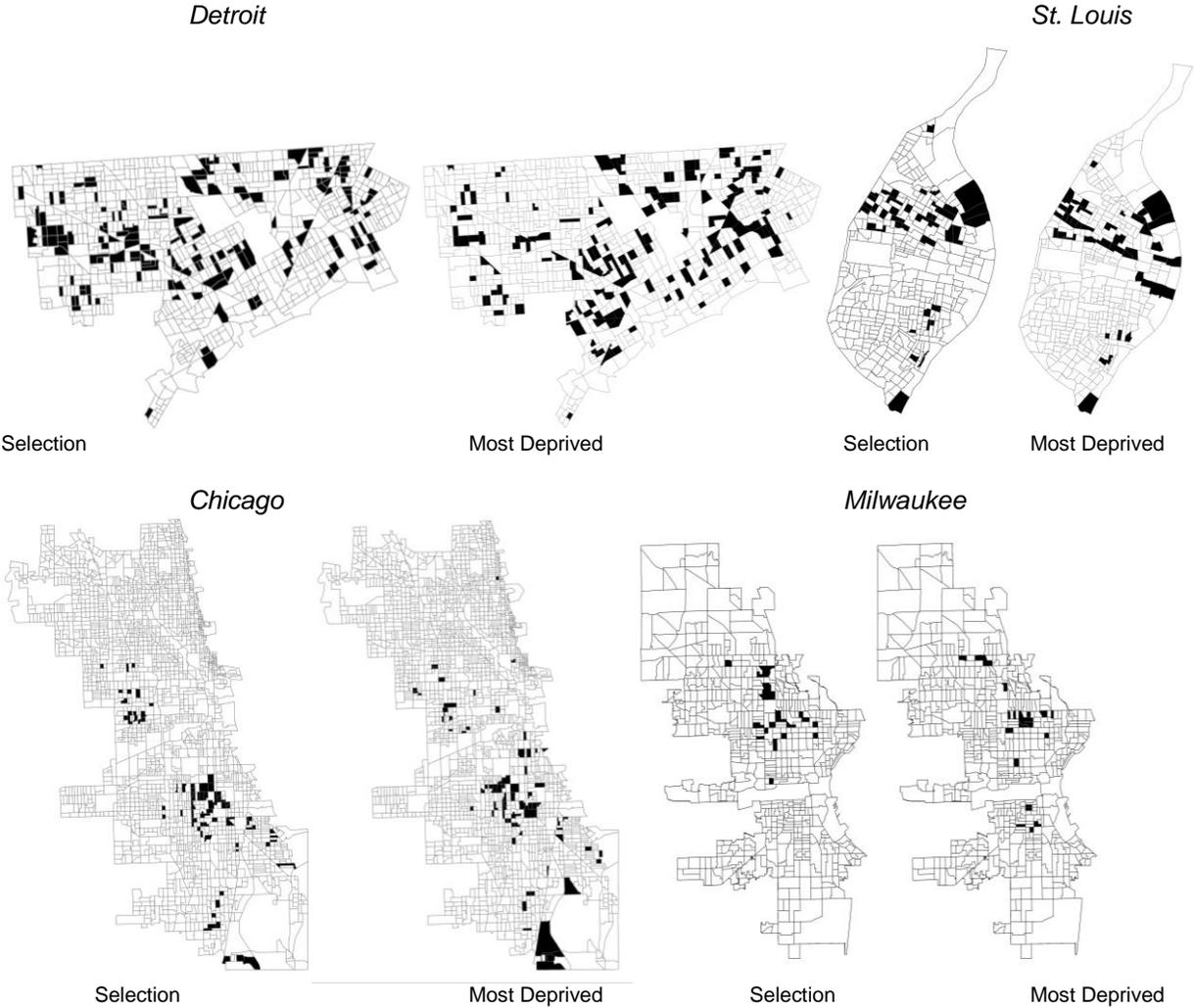

**Table 5. Median Values of Socioeconomic Indicators, City and Selected Areas.**

| Variable | Detroit City | Selected | Chicago City | Selected | Milwaukee City | Selected | Minneapolis City | Selected | St. Louis City | Selected |
|---|---|---|---|---|---|---|---|---|---|---|
| SD4DET | 7.783 | 9.493 | 4.454 | 9.816 | 4.749 | 10.218 | 2.346 | 15.617 | 5.295 | 9.602 |
| ABRPOP | 0.061 | 0.088 | 0.023 | 0.096 | 0.038 | 0.101 | 0.011 | 0.095 | 0.129 | 0.179 |
| MEDVAL | 41500 | 32550 | 199300 | 110950 | 113700 | 65600 | 181700 | 122500 | 100700 | 68800 |
| PERC40COMM | 15.949 | 15.827 | 36.364 | 43.678 | 10.280 | 15.294 | 9.189 | 27.368 | 11.832 | 17.755 |
| PERCBLACK | 95.797 | 96.073 | 11.190 | 98.305 | 21.163 | 90.890 | 10.867 | 0 | 47.144 | 97.780 |
| PERCHOUS30K | 35.471 | 42.529 | 0.000 | 5.472 | 0.000 | 0.000 | 0 | 49.222 | 3.409 | 14.425 |
| PERCPUBTRA | 7.299 | 10.000 | 24.764 | 35.963 | 6.291 | 15.789 | 11.075 | 39.815 | 8.036 | 20.623 |
| PERCRENT | 45.833 | 53.164 | 55.792 | 73.460 | 55.252 | 70.588 | 46.222 | 86.047 | 54.829 | 62.984 |
| PERCSNAP | 44.214 | 55.428 | 18.919 | 54.565 | 26.054 | 52.778 | 9.555 | 86.822 | 25.898 | 53.176 |
| PERCVAC | 30.233 | 42.745 | 11.525 | 35.864 | 8.303 | 35.135 | 6.644 | 38.278 | 18.234 | 39.541 |
| PERCWHITE | 2.968 | 2.669 | 51.802 | 0.272 | 55.507 | 3.289 | 75.874 | 4.086 | 43.169 | 1.937 |
| REVCOMM | 24.219 | 23.070 | 18.449 | 14.145 | 29.396 | 21.186 | 2.346 | 47.222 | 40.118 | 31.494 |
| N (% of total) | 868 | 156 (18) | 2152 | 71 (3.3) | 661 | 19 (2.9) | 378 | 1 (0.3) | 359 | 39 (10.9) |

**Table 6. Z-Scores By Which Detroit-Like Area Means Exceed Citywide Means.**

| | Detroit | Chicago | Milwaukee | St. Louis |
|---|---|---|---|---|
| SD4DET | -0.060 | 1.608* | 1.283* | 1.216* |
| ABRPOP | 0.329 | 0.394 | 1.746* | 0.430 |
| MEDVAL | -0.134 | -0.850 | -0.891 | -0.571 |
| PERC40COMM | 0.115 | 0.409 | 0.729 | 0.423 |
| PERCBLACK | -0.116 | 1.342* | 1.319* | 0.952 |
| PERCHOUS30K | 0.054 | 0.933 | -0.082 | 0.708 |
| PERCPUBTRA | 0.064 | 0.644 | 0.920 | 0.814 |
| PERCRENT | -0.015 | 0.796 | 0.718 | 0.441 |
| PERCSNAP | -0.130 | 1.826* | 1.361* | 1.361* |
| PERCVAC | -0.154 | 2.187** | 2.693** | 1.420* |
| PERCWHITE | -0.514 | -1.204* | -1.257* | -0.910 |
| REVCOMM | 0.082 | -0.373 | -0.292 | -0.332 |

Note: * = exceeds 1 standard deviation; ** = exceeds 2 standard deviations.

**Table 7. Median Values of Socioeconomic Indicators, Most Deprived Areas.**

| Variable | Detroit 10% (N=87) | N=156 | Chicago 10% (N=215) | N=71 | Milwaukee 10% (N=66) | N=19 | St. Louis 10% (N=36) | N=39 |
|---|---|---|---|---|---|---|---|---|
| SD4DET | 12.299 | 11.433 | 10.205 | 11.731 | 11.333 | 12.244 | 11.118 | 11.072 |
| ABRPOP | 0.065 | 0.066 | 0.075 | 0.074 | 0.057 | 0.063 | 0.171 | 0.172 |
| MEDVAL | 27900 | 30800 | 113400 | 102700 | 72000 | 70000 | 74950 | 73300 |
| PERC40COMM | 18.868 | 19.151 | 42.007 | 43.137 | 15.645 | 12.698 | 27.686 | 27.778 |
| PERCBLACK | 96.211 | 95.743 | 96.028 | 96.500 | 82.749 | 87.835 | 98.690 | 98.683 |
| PERCHOUS30K | 47.413 | 46.111 | 0.000 | 0.000 | 0.000 | 0.000 | 17.606 | 16.788 |
| PERCPUBTRA | 10.744 | 11.000 | 30.952 | 35.593 | 15.577 | 19.149 | 24.673 | 24.885 |
| PERCRENT | 60.221 | 56.039 | 73.725 | 79.741 | 73.889 | 76.889 | 69.053 | 68.000 |
| PERCSNAP | 64.972 | 60.678 | 54.483 | 62.542 | 63.495 | 68.240 | 57.784 | 56.637 |
| PERCVAC | 37.344 | 36.964 | 22.562 | 25.731 | 19.104 | 18.471 | 30.092 | 30.124 |
| PERCWHITE | 2.660 | 2.902 | 1.515 | 0.501 | 9.332 | 4.836 | 0.924 | 1.014 |
| REVCOMM | 26.364 | 25.168 | 19.672 | 17.890 | 25.323 | 25.420 | 35.282 | 37.063 |

**Table 8. Areas with Higher or Lower* Median Values.**

| Variable | Detroit | Chicago | Milwaukee | St. Louis |
|---|---|---|---|---|
| SD4DET | Deprivation | Deprivation | Deprivation | Deprivation |
| ABRPOP | Selection | Selection | Selection | Selection |
| MEDVAL* | Deprivation | Deprivation | Selection | Selection |
| PERC40COMM | Deprivation | Selection | Selection | Deprivation |
| PERCBLACK | Selection | Selection | Selection | Deprivation |
| PERCHOUS30K | Deprivation | Selection | Selection | Deprivation |
| PERCPUBTRA | Deprivation | Selection | Deprivation | Deprivation |
| PERCRENT | Deprivation | Deprivation | Deprivation | Deprivation |
| PERCSNAP | Deprivation | Deprivation | Deprivation | Deprivation |
| PERCVAC | Selection | Selection | Selection | Selection |
| PERCWHITE* | Selection | Selection | Selection | Deprivation |
| REVCOMM | Deprivation | Deprivation | Deprivation | Deprivation |